\documentclass{lmcs}

\keywords{Constraint Satisfaction, Universal Algebra, Deterministic Logspace}

\usepackage[numbers,square,comma]{natbib}
\usepackage{float}
\usepackage[pdfpagelabels]{hyperref}
\usepackage{amsmath}
\usepackage{amssymb}
\usepackage{amsthm}
\usepackage{graphicx}
\usepackage{geometry}
\usepackage{chngcntr}
\usepackage{thmtools}
\usepackage{thm-restate}
\usepackage{tabularx}
\usepackage{environ}

\counterwithin{figure}{section}

\theoremstyle{plain}\newtheorem{claim}[thm]{Claim}

\newcommand{\A}{\mathbb{A}}
\newcommand{\B}{\mathbb{B}}
\newcommand{\C}{\mathbb{C}}
\newcommand{\D}{\mathbb{D}}
\newcommand{\CSP}{\operatorname{CSP}}
\renewcommand{\t}{\theta}
\newcommand{\Sg}{\operatorname{Sg}}

\begin{document}
\title[CSPs over Multisorted Cores]{The Constraint Satisfaction Problem Over Multisorted Cores}

\author[D.~Delic]{Dejan Delic}
\author[J.~Marcoux]{John Marcoux}

\address{Department of Mathematics, Toronto Metropolitan University, Toronto, ON, M5B 2K3, Canada.}
\email{\{ddelic,jmarcoux\}@torontomu.ca}

\begin{abstract}
    Constraint Satisfaction Problems (CSPs, for short) make up a class of problems with applications in many areas of computer science. The first classification of these problems was given by Schaeffer who showed that every CSP over the domain \{0,1\} is either in \textsc{P} or is \textsc{NP}-complete. More recently this was shown to hold for all CSPs over finite relational structures independently by Bulatov and Zhuk. Furthermore, they characterized the complexity based solely on the polymorphism algebra of the associated relational structure, building upon the deep connections between universal algebra and complexity theory.

    In this article we extend this and consider what happens if the instance forms a special type of relational core called a \emph{multisorted core}. Our main result is that in this case the problem is reducible to computing the determinant of an integer valued matrix which places it in the complexity class \textsc{DET}, which is likely a strict subset of \textsc{P}. 
\end{abstract}

\maketitle

\section{Introduction}

The \emph{Constraint Satisfaction Problem} (CSP) is a fundamental object of study in Computer Science. A CSP instance asks if there is a mapping from a set of variables to a set of domains which satisfies some relation on the domains. Examples include the graph homomorphism problem, the graph colouring problem, scheduling problems, and the graph isomorphism problem. These problems have found applications in many areas of Computer Science including artificial intelligence~\cite{kolaitis2007logical}.

More formally, an \emph{instance} of the CSP is a triple $\mathcal{P}=(V,A,\mathcal{C})$, where $V=\{x_1,\ldots,x_n\}$ is a finite set of \emph{variables}, $A$ is a finite domain for the variables in $V$, and $\mathcal{C}$ is a finite set of \emph{constraints} of the form $C=(S,R_S)$, where $S$, the \emph{scope} of the constraint, is a $k$-tuple of variables $(x_{i_1},\ldots,x_{i_k})\in V^k$ and $R_S$ is a $k$-ary relation $R_S\subseteq A^k$, called the \emph{constraint relation} of $C$. A \emph{solution} for the instance $\mathcal{P}$ is any assignment $f:V\rightarrow A$ such that, for every constraint $C=(S,R_S)$ in $\mathcal{C}$, $f(S)\in R_S$. As an example, consider the {\sc $3$-Colouring} problem. {$3$-Colouring} asks if, given a graph $G$, it is possible to assign one of three colours to each vertex in $G$ such that if $u$ is adjacent to $v$, then $u$ and $v$ receive different colours. In this case the CSP has variables $V(G)$, domain $\{0,1,2\}$, and for each pair of adjacent vertices we have the constraint $\{(x,y) \mid x \neq y\}$.

A common approach to studying the complexity of the CSP is to restrict the relation on the domain to some fixed set of relations (as in the case of {\sc $3$-Colouring}). In~\cite{sch} Schaeffer used this approach to completely characterize the complexity of such CSPs over the domain $\{0,1\}$ as polynomial time or NP-complete. More recently Bulatov~\cite{bulatov2017dichotomy} and Zhuk~\cite{zhuk2020proof} independently proved the CSP dichotomy theorem, showing that any CSP where the relations come from a fixed template is either solvable in polynomial time or NP-complete.

From this point on, we will usually assume that the set of variables $V$ for an instance is an initial segment of the set of positive integers, i.e. $V=\{1,2,\ldots,n\}$, for some $n\geq 1$. A relational structure $\mathbf{A}=(A, \Gamma)$, defined over the domain $A$ of the instance $\mathcal{P}$, where $\Gamma$ is a finite set of relations on $A$, is often referred to as a \emph{relational template}, and the relations from $\Gamma$ form the signature of $\mathbf{A}$. An instance of $\CSP (\mathbf{A})$ will be an instance of the CSP such that all constraint relations belong to $\mathbf{A}$. 

The key tool in the study of these CSPs is the so-called \emph{algebraic approach} in which the algebra formed by certain types of symmetries of the relational template known as \emph{polymorphisms}. For example, the CSP dichotomy theorem states that the decision problem $\CSP(\mathbf{A})$ is solvable in polynomial time if and only if the associated algebra has a weak near-unanimity operation~\cite{bulatov2017dichotomy, zhuk2020proof}. We will give more details on the algebraic approach in Section~\ref{sec:homs}.

Our contribution is that if our CSP instance satisfies a certain core condition, then we can strengthen the statement from the dichotomy theorem to say that these CSPs are in a complexity class related to the complexity of computing matrix determinants which is likely a strict subset of P.

\begin{restatable*}{thm}{mainthm}
\label{thm:main}
For any finite relational template $\mathbf{A}$ and instance $\mathcal{P}$ of $\CSP (\mathbf{A})$, if $\mathcal{P}$'s corresponding multisorted structure is a core then $\mathcal{P}$ can be solved using a logspace Turing machine with access to an oracle for the class $\textsc{MOD}_k\textsc{L}$.
\end{restatable*}

\section{Background}

    In this section we give all the necessary background in order to prove the main result of Section $3$.

\subsection{Complexity}

    In this article, we will be primarily concerned with problems which can be placed in the computational complexity class \textsc{L} (deterministic logspace) and the functions which can be computed using Turing machines operating in deterministic logspace and using oracles of a counting nature, which are intimately related to the computational complexity of common problems in linear algebra over a finite ring.

A typical example of a problem which can be solved in deterministic logspace is the problem $(s,t)$-\textsc{UCONN} (connectivity on undirected graphs), which, given two vertices $s$ and $t$ in the input graph, asks whether there exists a path between $s$ and $t$. This rather deep fact will be used in the analysis of the complexity of the algorithm presented here and is due to Reingold~\cite{reingold2008undirected}.

A \emph{logspace transducer} is a Turing machine with a read-only input tape, a write-only
output tape, and a work tape which can contain at most $\mathcal{O}(\log n)$ symbols at any time. For that reason, one can view a logspace transducer as a function $F$ mapping instances of an algorithmic problem $\mathfrak{P}_1$ into instances of an algorithmic problem $\mathfrak{P}_2$ so that, if $\mathcal{I}$ is an instance of the problem $\mathfrak{P}_1$, the Turing machine for the function $F$, computing the instance $\mathcal{J}=F(\mathcal{I})$ of $\mathfrak{P}_2$, operates in deterministic logspace. It is a fairly elementary fact from the theory of computational complexity that a composition of any constant number of logspace transducers is again a logspace transducer. For a proof, see e.g. \cite{arora2009computational}.

In our algorithm, logspace transducers play an important role. The algorithm will be based on a series of reductions from one instance to another and those reductions will be carried out by logspace transducers which have access to a particular type of oracle. This will place the complexity of our algorithm in a particular complexity class, closely connected to the complexity of problems in linear algebra over finite rings $\mathbb{Z}_k$, $(k\in\mathbb{Z})$.

To explain this connection, we need to introduce the notion of particular subclasses of logspace with counting, the classes $\textsc{MOD}_k\textsc{L}$. The complexity class \#\textsc{L} consists of all computable functions $f:\{0,1\}^*\rightarrow\mathbb{N}$, such that there is a nondeterministic Turing machine $M$ using $\mathcal{O}(\log n)$ space, which halts on every input and along every computation path, so that the number of accepting paths on input $x$ is $f(x)$. For an integer $k\geq 2$, the class $\textsc{MOD}_k\textsc{ L}$ is defined to be the class of sets $A$ for which a function $f(x)\in \#$ \textsc{L} exists, such that, $x\in A$ if, and only if $f(x)\not \equiv 0 \mod k$. On the other hand, the complexity class \textsc{DET} consists of all problems that are $\textsc{NC}^1$-reducible to the problem of computing a determinant with entries from $\mathbb{Z}$, the ring of integers. We will not present a full definition of $\textsc{NC}^1$-reductions here but it would suffice to say that every reduction, which can be carried out by a logspace transducer is an $\textsc{NC}^1$-reduction. This is a direct consequence of the result from complexity theory that $\textsc{L}\subseteq\textsc{NC}^1$ (see e.g. \cite{arora2009computational}). We can consider proper subclasses of \textsc{DET}, $\textsc{DET}_k$, for every $k\geq 2$, which consist of all the problems, which can be $\textsc{NC}^1$-reduced to the problem of computing the determinant of a matrix over $\mathbb{Z}_k$.

Typical examples of problems which are complete for the class \textsc{DET} are the standard problems in linear algebra over $\mathbb{Z}$: rank computations, computing the determinant of a matrix, computing a solution of a linear system, computing a basis of a kernel of a linear transformation, computing the inverse of a matrix, and so on. For the subclasses $\textsc{DET}_k$, all these problems are still complete problems when relativized to matrices over $\mathbb{Z}_k$.

To prove our result, we will need the following theorem connecting \textsc{L} and $\textsc{MOD}_k\textsc{ L}$ to $\textsc{DET}_k$:

\begin{thmC}[\cite{buntrock1992structure}] For $k\geq 2$,

$$\textsc{L}^{\textsc{MOD}_k\textsc{ L}}\subseteq \textsc{DET}_k.$$
\end{thmC}

That is, every algorithmic problem which can be solved in logspace with the use of oracles which are in $\textsc{MOD}_k\textsc{ L}$ can be reduced to the problem of computing the determinant of a matrix over $\mathbb{Z}_k$. For further reading on computational complexity we direct the reader to~\cite{arora2009computational}. For more information on \textsc{MOD}-logspace classes and their properties, the reader is referred to \cite{buntrock1992structure} and \cite{hertrampf2000note}.

    \subsection{Algebras and Polymorphisms}\label{sec:homs}
\noindent In order to be able to fully utilize the power of the algebraic approach for studying the complexity of CSPs, we will now outline the connection between the constraint satisfaction problems on finite relational templates and their algebraic parametrization. We begin with the notion of a homomorphism between two relational structures.

An \emph{$n$-ary operation} or an \emph{operation of arity $n$} on a set $A$ is a mapping of the form $f:A^n\rightarrow A$. Let $f$ be an $n$-ary operation on $A$ and let $k>0$. We write $f^{(k)}$ to denote the $n$-ary operation obtained by applying $f$ coordinate-wise on $A^k$. That is, we define the $n$-ary operation $f^{(k)}$ on $A^k$ by
\[
f^{(k)}(\mathbf a^1,\dots,\mathbf a^n)=(f(a^1_1,\dots,a^n_1),\dots,f(a^1_k,\dots,a^n_k)),
\]
for $\mathbf a^1,\dots, \mathbf a^n\in A^k$.

  Let $\mathbf A$ and $\mathbf B$ be relational structures in the same signature~$\Gamma$. A \emph{homomorphism} from $\mathbf A$ to $\mathbf B$ is a mapping $\varphi$ from $A$ to $B$ such that for each $k$-ary relation symbol $R$ in $\Gamma$ and each $k$-tuple $\mathbf{a}\in A^k$, if $\mathbf{a}\in R^\mathbf A$, then $\varphi^{(k)}(\mathbf{a})\in R^\mathbf B$. We write $\varphi:\mathbf A\to\mathbf B$ to mean that $\varphi$ is a homomorphism from $\mathbf A$ to $\mathbf B$, and $\mathbf A\to\mathbf B$ to mean that there exists a homomorphism from $\mathbf A$ to $\mathbf B$. An \emph{isomorphism} is a bijective homomorphism $\varphi$ such that $\varphi^{-1}$ is also a homomorphism. A homomorphism $\mathbf A\to\mathbf A$ is called an \emph{endomorphism}. 

A finite relational structure $\mathbf A'$ is a \emph{core} if every endomorphism of $\mathbf A'$ is surjective. For every $\mathbf A$ there exists a relational structure $\mathbf A'$ such that $\mathbf A\to\mathbf A'$ and $\mathbf
A'\to\mathbf A$ and $\mathbf A'$ is of minimum size with respect to these properties; that structure $\mathbf A'$ is called the \emph{core of $\mathbf A$}. The core of $\mathbf A$ is unique (up to isomorphism) and $\CSP(\mathbf A)$ and $\CSP(\mathbf A')$ are equivalent decision problems. Equivalently, the core of $\mathbf A$ can be defined as an induced substructure of minimum size that $\mathbf A$ retracts onto. (See~\cite{helnes} for details on cores for graphs, cores for relational structures are a natural generalization.)
    
We now define some concepts from universal algebra which we will be needed to define the polymorphism algebra of a relational structure. An \emph{algebra} is an ordered pair $\mathbb{A}=(A, F)$, where $A$ is a nonempty set, the \emph{universe} of $\mathbb{A}$, while $F$ is the set of \emph{basic operations} of $\mathbb{A}$, consisting of functions of arbitrary, but finite, arities on $A$. The list of function symbols and their arities is the \emph{signature} of $\mathbb{A}$. 

    A \emph{subuniverse} of an algebra $\mathbb{A}$ is a nonempty subset $B\subseteq A$ closed under all operations of $\mathbb{A}$. If $B$ is a subuniverse of $\mathbb{A}$, by restricting all operations of $\mathbb{A}$ to $B$, such a subuniverse is a \emph{subalgebra} of $\mathbb{A}$, which we denote $\mathbb{B}\leq \mathbb{A}$. In particular we will be interested in subalgebras generated by a two element subset of $A$. For two distinct elements $a,b \in A$ we use $\Sg(a,b)$ to refer to the smallest subalgebra of $\A$ which contains both $a$ and $b$.

If $\{\mathbb{A}_i\}_{i \in I}$ is an indexed family of algebras of the same signature, the product $\prod_{i \in I} \mathbb{A}_i$ of the family is the algebra whose universe is the Cartesian products of their universes endowed with the basic operations which are coordinate-wise products of the corresponding operations in $\mathbb{A}_i$. If $\mathbb{A}$ is an algebra, its $n$-th power will be denoted $\mathbb{A}^n$.

An equivalence relation $\alpha$ on the universe $A$ of an algebra $\mathbb{A}$ is a \emph{congruence} of $\mathbb{A}$, if $\alpha \leq \mathbb{A}^2$, that is to say, if $\alpha$ is preserved by all basic operations of $\mathbb{A}$. In that case, one can define the algebra $\mathbb{A}/\alpha$, the \emph{quotient of} $\mathbb{A}$ \emph{by} $ \alpha$, with the universe consisting of all equivalence classes (cosets) in $A/\alpha$ and whose basic operations are induced by the basic operations of $\mathbb{A}$. The $\alpha$-congruence class containing $a\in A$ will be denoted $a/\alpha$.

An algebra $\mathbb{A}$ is said to be \emph{simple} if its only congruences are the trivial, diagonal relation $\{(a,a)\mid a\in A\}$ denoted $0_{\mathbb{A}}$ and the full relation, $\{ (a,b)\mid a,b\in A\}$ denoted $1_{\mathbb{A}}$. It is a well-known fact (see e.g. \cite{Burris1981}) that the congruences of $\mathbb{A}$ form a lattice $Con(\mathbb{A})$; namely, for any $\alpha,\beta\in Con(\mathbb{A})$, $\alpha\wedge\beta$ is the intersection of $\alpha$ and $\beta$, while $\alpha\vee \beta$ is the smallest congruence containing both $\alpha$ and $\beta$.

Any subalgebra of a Cartesian product of algebras $\mathbb{A}\leq \prod_i \mathbb{A}_{i\in I}$ is equipped with a family of congruences arising from projections on the product coordinates. We denote $\pi_i$ the congruence obtained by identifying the tuples in $A$ which have the same value in the $i$-th coordinate. Given any $J\subseteq I$, we can define a subalgebra of $\mathbb{A}$, $proj_J(\mathbb{A})$, which consists of the projections of all tuples in $A$ to the coordinates from $J$. If $\mathbb{A}\leq \prod_{i\in I} \mathbb{A}_i$ is such that $proj_i (\mathbb{A})=\mathbb{A}_i$, for every $i\in I$, we say that $\mathbb{A}$ is a \emph{subdirect product} and denote this fact $\mathbb{A}\leq_{sp} \prod_{i\in I} \mathbb{A}_i$.

If $\mathbb{A}$ and $\mathbb{B}$ are two algebras of the same signature, a mapping from $A$ to $B$ which preserves all basic operations is a \emph{homomorphism}. An \emph{isomorphism} is a bijective homomorphism between two algebras of the same signature.

Given an algebra $\mathbb{A}$, a \emph{term} is a syntactical object describing a composition of basic operations of $\mathbb{A}$. A \emph{term operation} $t^\mathbb{A}$ of $\mathbb{A}$ is the interpretation of the syntactical term $t(x_1,\ldots,x_m)$ as an $m$-ary operation on $A$, according to the formation tree of $t$. A \emph{variety} is a class of algebras of the same signature, which is closed under the class operators of taking products, subalgebras, and homomorphic images (or, equivalently, under the formation of quotients by congruence relations.) The variety $\mathcal{V}(\mathbb{A})$ generated by the algebra $\mathbb{A}$ is the smallest variety containing $\mathbb{A}$. Birkhoff's theorem (see \cite{Burris1981}) states that every variety is an equational class; that is, every variety $\mathcal{V}$ is uniquely determined by a set of identities (equalities of terms) $s\approx t$ so that $\mathbb{A}\in\mathcal{V}$ if and only if $\mathbb{A}\models s\approx t$, for every identity $s\approx t$ in the set. 
  
    We are now ready to define the \emph{polymorphism algebra} associated with a relation template $\mathbf{A}$. Polymorphisms are a natural generalization of endomorphisms to higher arity operations. Given a $\Gamma$-structure $\mathbf{A}$, an $n$-ary \emph{polymorphism} of $\mathbf{A}$ is an $n$-ary operation $f$ on $A$ such that $f$ preserves the relations of $\mathbf A$. That is, if $\mathbf{a}^1,\dots,\mathbf{a}^n\in R$, for some $k$-ary relation $R$ in $\Gamma$, then $f^{(k)}(\mathbf a^1,\dots,\mathbf a^n)\in R$. Furthermore, if a relational structure $\mathbf{A}$ is a core, one can construct a structure $\mathbf{A}'$ from $\mathbf{A}$ by adding, for each element $a\in A$, a unary constraint relation $\{a\}$. This enables us to further restrict the algebra of polymorphisms associated with the template; namely, if $f(x_1,\ldots,x_m)$ is an $m$-ary polymorphism of $\mathbf{A}'$, it is easy to see that $f(a,a,\ldots,a)=a,$ for all $a\in A$. In addition to this, the constraint satisfaction problems with the templates $\mathbf{A}$ and $\mathbf{A}'$ are log-space equivalent. Therefore, we may assume that the algebra of polymorphisms associated to any CSP under consideration is \emph{idempotent}; i.e. all its basic operations $f$ satisfy the identity
$$f(x,x,\ldots,x)\approx x.$$
We note here that the idea of a polymorphism algebra is the primary tool for studying CSPs and motivates the introduction of the algebraic tools we introduce in the remainder of this section.

A ternary operation $m:A^3\rightarrow A$ on a finite set is said to be \emph{Maltsev} if it satisfies the algebraic identities $m(x,x,y)\approx m(y,x,x)\approx y$. An algebra $\mathbb{A}=(A, F)$ is said to be a \emph{Maltsev algebra}, if its set of operations $F$ contains a Maltsev operation. In the construction of the algorithm which will be the central result of the paper, a particular type of Maltsev algebras will play a fundamental role. These are \emph{affine} algebras, which can be viewed as algebras defined from Abelian $p$-groups $\mathbb{A}=(A, \{+,0\})$, where $+$ is the usual addition in the arithmetic modulo $p$. It is easily seen that any such algebra $\mathbb{A}$ is a Maltsev algebra by considering the ternary operation $m(x,y,z)=x-y+z$. Clearly, any Maltsev operation $m(x,y,z):A^3\rightarrow A$ is idempotent, since it satisfies $m(x,x,x)\approx x$, for all $x\in A$. Therefore, every singleton $\{x\}$ is a subalgebra of $\mathbb{A}=(A;\{m(x,y,z)\})$. A typical example of a constraint satisfaction problem over a finite Maltsev template is the problem of solving a system of linear equations in $n$ variables over a fixed finite field $K$. Let $S$ be its solution space viewed as an $n$-ary relation on $K$. The operation $m(x,y,z)=x-y+z$ is a polymorphism of the relational structure $\mathbf{S}=(K; S)$. The converse is also true; namely, one can show that any $n$-ary relation on $K$, for $n\geq 1$, which has $m(x,y,z)$ as its polymorphism is a solution of some system of linear equations over $K$ in $n$ variables.

Finally, we state some facts about subdirect products of Maltsev algebras which will be needed later. The first fact concerns the connectivity in subdirect products of simple Maltsev algebras as bipartite graphs (i.e. for an algebra $\mathbb{C} \leq_{sp} \mathbb{A} \times \mathbb{B}$ the graph with vertex set $A \cup B$ and edges of the form $xy$ if and only if $xy \in C$). For the proof, see e.g. \cite{Burris1981}.

\begin{thm} \label{maltsev} Let $\mathbb{A}_1,\ldots,\mathbb{A}_n$ be simple algebras in a Maltsev variety. If
\[\mathbb{B}\leq_{sp} \mathbb{A}_1\times\ldots\times\mathbb{A}_n\]
is a subdirect product, then
\[\mathbb{B}\cong \mathbb{A}_{i_1}\times\ldots\times\mathbb{A}_{i_k}\]
for some $\{i_1,\ldots,i_k\}\subseteq \{1,\ldots,n\}$. 

In particular, if $\mathbb{A}$ and $\mathbb{B}$ are two simple Maltsev algebras then any subdirect product $\mathbb{C}\leq_{sp} \mathbb{A}\times\mathbb{B}$ is either the direct product or the graph of an isomorphism $f:\mathbb{A}\rightarrow\mathbb{B}$.
\end{thm}

The existence of a Maltsev operation implies the following property on any subdirect product, which we will refer to as the \emph{rectangularity property}.

\begin{prop}\label{rect} 
    Let $C\leq_{sp} \mathbb{A}\times\mathbb{B}$, where $\mathbb{A}$ and $\mathbb{B}$ are algebras in a Maltsev variety. Then, the following holds: if $(a,b), (a,b'), (a',b')\in C$, then $(a',b)\in C$.
\end{prop}

    \proof Let $m(x,y,z)$ be a Maltsev operation on both $\mathbb{A}$ and $\mathbb{B}$. Then,
$$(a',b)=(m(a,a,a'), m(b,b',b'))\in C.$$\qedhere

Using Proposition \ref{rect}, one can prove a generalization of the second part of Theorem \ref{maltsev}.

\begin{thm} \label{link} 
    If $\mathbb{A}$ and $\mathbb{B}$ are two Maltsev algebras, with $\mathbb{B}$ being simple, then any subdirect product $\mathbb{C}\leq_{sp} \mathbb{A}\times\mathbb{B}$ is either the direct product or, there exists a maximal congruence $\theta\in\operatorname{Con}(\mathbb{A})$, such that $C$ is the graph of an isomorphism $f:\mathbb{A}/\theta\rightarrow\mathbb{B}$. In particular, the congruence $\theta$ is defined as follows: for $a,a'\in A$, $(a,a')\in \theta$ if, and only if, there exists $ b\in B $ such that $(a,b),(a',b)\in C.$
\end{thm}

In the first case in the statement of Theorem \ref{link}, we refer to the subdirect product as \emph{linked} while, in the second case, we will say that $C$ is \emph{unlinked}. For further reading on universal algebra we direct the reader to~\cite{Burris1981}.

Finally, recall that an algebra is said to be \emph{Taylor} if it has an $n$-ary ($n \geq 2$) operation $t$ satisfying\[
    t(y,x,x,\dots,x,x) \approx t(x,y,x,\dots,x,x) \approx \cdots \approx t(x,x,x,\dots,x,y)
\]

We will be focusing on Taylor algebras for the remainder of the paper as they are precisely the algebras whose corresponding CSPs are tractable~\cite{bulatov2017dichotomy,zhuk2020proof}.

\subsection{Idempotent Algebras Viewed as Graphs with Coloured Edges}

We now review the concept of the coloured graph of an algebra as introduced by Bulatov in~\cite{bulatov2004graph}. While this approach can also be thought of through the language of tame congruence theory~\cite{hobbymckenzie}, we use the language introduced by Bulatov as it still has enough expressive power for this paper.

First recall that a binary operation $f$ is said to be \emph{semilattice} if it satisfies $f(x,x) = x$, $f(x,y)=f(y,x)$, and $f(x,f(y,z)) = f(f(x,y),z)$. A ternary operation $g$ is said to be \emph{majority} if it satisfies $g(x,x,y) = g(x,y,x) = g(y,x,x) = x$. Finally, a ternary operation $m$ is \emph{affine} if it is Maltsev and the algebra is a module.

Let $\mathbb{A}$ be a finite idempotent algebra. A pair of elements $\{a,b\}$ will be called an \emph{edge}, if there exists a congruence $\theta$ of $\operatorname{Sg}(a,b)$ and a term operation $f$ of $\mathbb{A}$ such that, $f$ induces either

\begin{itemize}
\item[(a)] a binary semilattice operation on $\operatorname{Sg}(a,b)/\theta$; or
\item[(b)] a ternary majority operation on $\operatorname{Sg}(a,b)/\theta$; or
\item[(c)] a ternary affine operation on $\operatorname{Sg}(a,b)/\theta$.
\end{itemize} 

An edge $\{a,b\}$ is said to be \emph{thin} if $\theta$ is the identity relation on $\operatorname{Sg}(a,b)$. We define the colouring of all edges in a finite idempotent algebra $\mathbb{A}$ in the following way: if an edge is of the semilattice type, it will be coloured red; if an edge admits a ternary majority operation, it will be assigned the colour yellow; and, if it is of the affine type, it will be coloured blue. If an algebra has a Taylor term or, equivalently, if its associated constraint satisfaction problem is not \textsc{NP}-complete, all of its edges will be assigned one of the three colours. The undirected graph $\operatorname{Gr}(\mathbb{A})$, whose vertices are elements of the algebra and whose edge relation is defined as above, will also be connected~\cite{bulatov2004graph}.

\subsection{Reduction of General CSPs to Binary Relational Structures}

In this subsection, we outline a reduction of an instance $\mathcal{I}$ of $\operatorname{CSP}(\mathbf{A})$, where $\mathbb{A}$ is a finite idempotent algebra, to a binary instance over a binary relational template, parametrized by $\mathbb{A}^m$, for some $m\geq 1$. The construction is due to L. Barto and M. Kozik and we largely adhere to their exposition in~\cite{b-k1}. The reduction is given by first-order (in fact, quantifier-free) formulas in a bounded number of variables, and can be carried out in deterministic logspace.

An instance $\mathcal{P}$ is said to be \emph{syntactically simple} if it satisfies the following conditions:

\begin{itemize}
\item Every constraint in $\mathcal{C}$ is binary and its scope is a pair of variables $(x,y)\in V^2$.
\item For every pair of variables $x,y$, there is precisely one constraint $E_{x,y}$ with the scope $(x,y)$.
\item If $x=y$, then $E_{x,x}=\{(a,a)\,\vert\, a\in P_x\}$, where $P_x$ is the $x$-th domain.
\item If $(x,y)$ is the scope of $E_{x,y}$, then $(y,x)$ is the scope of the constraint $E_{y,x}=\{(b,a) \, \vert\, (a,b)\in E_{x,y}\}$ (\emph{symmetry of constraints}).
\end{itemize}

Given any finite algebra $\mathbb{A}$ parameterizing the instance $\mathcal{I}$ such that the maximal arity of a relation in $\mathcal{I}$ is $p$, we define a new, syntactically simple instance $\mathcal{P}$ in the following way:

\begin{itemize} 
\item The instance is parametrized by $\mathbb{A}^{\lceil \frac{p}{2}\rceil}$, which is an algebra satisfying all term identities (equations) $s\approx t$, satisfied by $\mathbb{A}$. 
\item For every $\lceil\frac{p}{2}\rceil$-tuple of variables in $\mathcal{I}$, we introduce a new variable in $\mathcal{P}$ and, if $x=(x_1,\ldots,x_{\lceil\frac{p}{2}\rceil})$ and $y=(y_1,\ldots,y_{\lceil\frac{p}{2}\rceil})$ with $x\neq y$, we introduce a constraint 
\begin{align*}
E_{x,y} &=\{((a_1,\ldots,a_{\lceil\frac{p}{2}\rceil}),(b_1,\ldots,b_{\lceil\frac{p}{2}\rceil}))\,\vert \, (a_1,\ldots,a_{\lceil\frac{p}{2}\rceil},b_1,\ldots,b_{\lceil\frac{p}{2}\rceil})\\
&\mbox{ is a $p$-assignment of values which satisfies all atomic formulas }\\
& \mbox{ on the tuples of variables } x,y \}
\end{align*}
while, if $x=y$, the relation $E_{x,x}$ is simply the equality of $\lceil\frac{p}{2}\rceil$-tuples in $\mathbb{A}^{\lceil\frac{p}{2}\rceil}$.
\end{itemize}

The binary instance $\mathcal{P}$ constructed in this way will have a solution if, and only if, the instance $\mathcal{I}$ has a solution. 

From the reduction described above, it is easily seen that, if $\mathcal{I}$ is an instance parametrized by a finite algebra $\mathbb{A}$, then the constructed, syntactically simple binary instance $\mathcal{P}$ can be parametrized by the direct product $\mathbb{A}^{\lceil\frac{p}{2}\rceil}$. In particular, if the original instance $\mathcal{I}$ is parametrized by a Maltsev algebra, then so is $\mathcal{P}$.

\subsection{Properties of Multisorted Cores}\label{sec:multisorted}

Based on the previous section, if $\mathcal{I}$ is an instance parametrized by an algebra $\mathbb{A}$, then the constructed, syntactically simple binary instance $\mathcal{P}$ can be parametrized by the direct product $\mathbb{A}^{\lceil\frac{p}{2}\rceil}$, where $p$ is the maximum arity of a relation in the associated template $\mathbf{A}$. In particular, if the original instance $\mathcal{I}$ is parametrized by an algebra having a Taylor term, so is $\mathcal{P}$. More specifically, if the new parameterizing algebra $\mathbb{A}^{\lceil\frac{p}{2}\rceil}$ is a Taylor algebra, then so is every one of its subalgebras.

Next, we will consider what happens if the multisorted structure (multiconsistency graph) $\mathbf{B}_{\mathcal{P}}$, corresponding to the instance $\mathcal{P}$ is a relational core. This will rely on the fact that the constraint satisfaction problem $\mathcal{P}$ and the graph homomorphism problem with the template graph $\mathbf{B}_{\mathcal{P}}$ are first-order equivalent as algorithmic problems. To see this simply note that any solution is a clique subgraph with one point in each domain and a homomorphism from a complete graph to $\mathbf{B}_{\mathcal{P}}$ must send each point to a distinct domain. Since first-order equivalence is weaker than equivalence under logspace reductions we can think of these problems as being the same for our purposes. We note here that since the size of the multisorted structure is dependent on the size of the instance, we can no longer recognize whether or not our template is a core in polynomial time. As such, for the remainder of the paper we will make the assumption that the instances we consider have their corresponding multisorted structure be a core.

 Let $\mathbb{B}$ be a subalgebra of some $\mathbb{P}_i$. We will say that a subalgebra $\mathbb{A}\leq \mathbb{B}$ is an \emph{absorbing subuniverse} of $\mathbb{B}$, if there exists an $m$-ary ($m\geq 2$) term (polymorphism) such, that for all $a\in A, b\in B$, the following holds
$$t(b,a\ldots,a,),t(a,b,a\ldots,a),\ldots,t(a,\ldots,a,b)\in A.$$
For example, the 3-ary majority polymorphism demonstrates that every singleton is an absorbing subuniverse. We will write that as $\mathbb{A} \: \triangleleft \: \mathbb{B}$. If an algebra has no proper absorbing subuniverses (other than itself), we say that it is \emph{absorption-free}.

If our instance, viewed as a multiconsistency graph, is a core, it cannot have any endomorphisms which are not surjective. Consider a finite subalgebra $\mathbb{B}$ such that one of the $\mathbb{P}_i$'s has a proper absorbing subuniverse $C \: \triangleleft \: \mathbb{B}$. We will show that no solution $f$ can be such that $f(i)\in C$.

\begin{claim} \label{claim:absorption}
    Let $C \: \triangleleft \: \mathbb{B}\leq \mathbb{P}_i$, for some $i\in V$, with $\mathbb{B}$ finite, and let this absorption be witnessed by an $m$-ary term $t$, where $m\geq 2$. Then, there is no solution $f: V\rightarrow A$, such that $f(i)\in C$.
\end{claim}

\proof We argue by contradiction. So, let us assume that such a solution $f:V\rightarrow A$ exists. Then, as in~\cite{berkholz2015limitations}, we define a homomorphism (endomorphism) which acts on all domains $\mathbb{P}_j$ in the following way:
$$s(x)=t(f(j),\ldots,f(j),x,f(j),\ldots,f(j)).$$
In particular, for our fixed $i\in V$, $\mathbb{C}$ absorbs $\mathbb{B}$, and are both finite subuniverses of $\mathbb{P}_i$. So, $s(x)$ will collapse at least two different elements of $\mathbb{B}$ (and, then, of $\mathbb{P}_i$) so the mapping $s(x)$ will fail to be surjective in at least one domain $\mathbb{P}_i$, which contradicts the fact that the instance is a core.\qed

As a consequence of Claim~\ref{claim:absorption}, given a two-generated subuniverse $\operatorname{Sg}_i(a,b)$ of any $\mathbb{P}_i$, where $\mathbb{P}_i$ denotes the $i$-th domain of $\mathbf{B}_{\mathcal{P}}$, no proper absorbing subuniverse of $\operatorname{Sg}_i(a,b)$ of any $\mathbb{P}_i$ can contain a solution. Also, we make note of the following fact: a proper absorbing subuniverse of $\operatorname{Sg}_i(a,b)$ cannot contain both $a$ and $b$, since, then, it would coincide with the entire $\operatorname{Sg}_i(a,b)$.

In particular, we make note of the following facts:

\begin{enumerate}
\item If $Sg_A(a,b)$ has a semilattice operation $f$ on $Sg_A(a,b)/\theta_{a,b}$ with the minimum element $c/\theta_{a,b}$, then no solution passing through $Sg_A(a,b)$ can pass through $c/\theta_{a,b}$.
\item If $Sg_A(a,b)$ has a majority operation $f$ on $Sg_A(a,b)/\theta_{a,b}$ with the minimum element $c/\theta_{a,b}$, then there is no solution passing through $Sg_A(a,b)$ since every element $c/\theta_{a,b}$ is an absorbing subuniverse for the operation $f$ in its own right.
\end{enumerate}

\subsection{Symmetric Linear Datalog and (1,2)-Consistency}

A \emph{Datalog program} for a relational template $\mathbf{A}$ is a finite set of rules of the form $T_0\leftarrow T_1,T_2,\ldots, T_n$ where the $T_i$'s are atomic formulas. 
$T_0$ is the \emph{head} of the rule, while $T_1,T_2,\ldots, T_n$ form the \emph{body} of the rule.
Each Datalog program consists of two kinds of relational predicates:
the \emph{intentional} ones (IDBs), which are those occurring at least once in the head of some rule, and which are not part of the original signature of the template (they are derived by the computation).
The remaining predicates are said to be the \emph{extensional} ones, or EDBs. They are relations from the signature of the template and do not change during computation; i.e. they cannot appear in the head of any rule.
In addition to those, there is one special, designated IDB, which is nullary (Boolean) and referred to as the \emph{goal} of the program. The semantics of Datalog programs are generally defined
in terms of fixed-point operators. 

 A rule $T_0\leftarrow T_1,T_2,\ldots, T_n$ is said to be \emph{linear} if at most one atomic formula in its body is an IDB. A Datalog program is said to be \emph{linear} if all its rules are linear. The evaluation of a linear Datalog program is in nondeterministic logspace since, from the computational complexity point of view, it reduces to repeated connectivity checks in a finite directed graph corresponding to the program. The linear rules in which an IDB appears in the body are said to be \emph{recursive}.

The \emph{symmetric complement} of a recursive linear rule $T_0\leftarrow T_1,T_2,\ldots, T_n$ in which, without loss of generality, the IDBs are $T_0$ and $T_1$, is defined to be the rule
$$T_1\leftarrow T_0,T_2,\ldots, T_n.$$ If the rule is non-recursive, its symmetric complement is the rule itself. A linear Datalog program is \emph{symmetric}, if the symmetric complement of every rule also appears in the program.

Given a CSP instance $\mathcal{I}=(V, A, \mathcal{C})$ over a relational template $\mathbf{A}$, its \emph{canonical Symmetric} (1,2)-\emph{Datalog program} has a unary IDB, $P_i(x)$, for each domain $P_i\subseteq A$ in the instance, with the said IDBs being the only IDBs in the program. The program allows derivation rules with the body involving at most two variables along with their symmetric complements. 1 indicates that the program is deriving facts about unary relations only, while $2$ indicates that the maximal number of distinct variables in any rule is 2. 

If the instance $\mathcal{I}$ is such that:

\begin{enumerate}
\item all constraints are binary; and
\item for every pair of variables $i,j\in V$, there exists a unique binary constraint $E_{i,j}\in \mathcal{C},$ so that $E_{i,j}=E_{j,i}^{-1}$,
\end{enumerate}

then if the canonical symmetric (1,2)-Datalog program does not derive a contradiction (empty instance), all constraints in the derived instance will be subdirect products. This is due to the fact that the only recursive rules involving unary IDBs $R_i, R'_i$ and $R_j, R'_j$, defined on $P_i$ and $P_j$, respectively, are of the form $R_j (x)\leftarrow E_{i,j}(y,x),R_i(y)$, $R'_j(x)\leftarrow E_{j,i}(x,y),R'_i(y)$, along with their symmetric complements.

Finally, the complexity of computing the canonical symmetric (1,2)-Datalog program for any (not necessarily binary) instance $\mathcal{I}$ of $\operatorname{CSP}(\mathbf{A})$ is in deterministic logspace~\cite{reingold2008undirected} since the run of the program can be viewed as a sequence of connectivity checks in an undirected graph, associated with $\mathcal{I}$. This fact will play an important role in the construction of a \textsc{DET} algorithm for solving instances of CSPs parametrized by finite Maltsev algebras.

\section{The Algorithm}

This section is divided into three parts which will constitute the proof of the below theorem. 

\mainthm

The first subsection outlines our algorithm and establishes its complexity. In the second subsection we provide the details for a reduction of our template based on the coloured graph of the corresponding algebra which allows us to assume all simple subalgebras are affine. The final subsection gives a short proof that our algorithm does in fact correctly decide the CSP. For the remainder of the section we will assume that all relational structures are finite multisorted cores whose associated algebras are Taylor.

\subsection{Outline of the Algorithm}

The main component of our algorithm is what we will call the $\#\textsc{L}$ consistency checks. In order to describe these we first define an \emph{isomorphism path} as the edges corresponding to a sequence of isomorphisms $\phi_1:D_1 \to D_2/\t_2, \phi_2: D_2/\t_2 \to D_3/\t_3, \dots \phi_{k-1}: D_{k-1}/\t_{k-1} \to D_k/\t_k$ as in Theorem~\ref{maltsev}. We will show that by considering the subinstances corresponding to the subalgebras $D_1/\t_1, D_2/\t_2,\dots,D_k/\t_k$, we will be able to decide whether or not the CSP is satisfiable.

We will show later that for any two-generated subalgebra $\operatorname{Sg}(a,b)$ the following holds. If it corresponds to a yellow edge, then we can eliminate both $a$ and $b$ from our domain, and if it corresponds to a red edge, then we can eliminate at least one of $a$ and $b$ from our domain. Thus from here on out we assume that all two-generated subalgebras correspond to blue edges. Furthermore, for any simple subalgebra corresponding to a yellow edge under some maximal congruence, we will get a yellow edge between two blocks of the congruence which implies the existence of a yellow edge in $\operatorname{Gr}(\mathbb{A})$. Thus we will assume that all simple subalgebras are affine.

Suppose we have a point $x$ and we wish to check it for $\#\textsc{L}$ consistency. First, we reduce all our domains as described above and run Datalog consistency in order to reduce to subdirect products. If $x$ is not eliminated by the Datalog check, fix a simple subalgebra $\B$ containing $x$. Take the quotient CSP formed by taking all isomorphic copies of $\B$ which are reachable along isomorphism type paths. That is to say, delete all vertices outside these domains and keep exactly one vertex from each equivalence class of the quotient relation which yields an isomorphism. Note that this can be found in logspace since undirected connectivity is in $\textsc{L}$. If, for some choice of $\B$, this quotient CSP has no solution then we reject $x$, otherwise $x$ passes the $\#\textsc{L}$ consistency check. Next, we demonstrate that solving the quotient CSP is equivalent to solving a system of equations over some finite field so this can be done in $\textsc{MOD}_k \textsc{L}$ for some $k$ depending on the template and choice of subalgebra. 

Namely, the domains of the quotient CSP are isomorphic copies of the same affine group $\mathbb{Z}_{q}$, where $q = p^t$ is a prime power. For such an instance, the restrictions of binary constraints of the original problem to the new domains are either (1) a graph of an isomorphism between two copies of the same affine group over $\mathbb{Z}_{q}$; or (2) a full direct product between two affine groups. The constraints of type (2) do not restrict the solution set of the quotient instance and can, therefore, be disregarded in solving it. Thus, we are left with the binary constraints of type (1). By fixing an element whose role is to act as the zero of the underlying affine group in each domain, every domain becomes an Abelian group with the gorup operation derived from the corresponding ternary affine operation. In that case, every restricted binary constraint of type (1) between two domains, which are isomorphic as affine structures, can be represented as a linear equation in two variables over the Abelian group $\mathbb{Z}_{q}$. Consequently, the quotient instance can be solved as a system of linear equations, all of them in two variables, over the Abelian group $\mathbb{Z}_{q}$, which places it in the complexity class $\textsc{MOD}_p \textsc{L}$. 

The number of subalgebras containing $x$ is bounded above by a constant so we can execute all these checks sequentially with only a constant amount of extra space required. Thus we can check the $\#\textsc{L}$ consistency of $x$ in $\textsc{MOD}_k \textsc{L}$ for some $k$ depending on all the choices of $p$ we were required to make earlier. Note as well that this can be viewed as a logspace algorithm with access to a bounded number of $\textsc{MOD}_k \textsc{L}$ oracles which places it in the complexity class $\textsc{DET}_k$ as well.

Thus to build our solution we can check each point one by one to see if they are consistent and if there are points in every domain which are not eliminated then we are guaranteed a solution by the proof of correctness. The algorithm we describe will be in $\textsc{MOD}_k \textsc{L}$ for some $k$ depending on all our required choices of $p$ which will depend only on the algebra, and the other parts of the algorithm run in deterministic logspace.

\subsection{Type Reduction}

We now prove the previously mentioned reductions. Under the assumption that $\mathbf{A}$ is a multisorted core, we first claim that the yellow edges in the colouring, must be absent.

Suppose $(a,b)$ is a yellow edge in the colouring of $\A$. Then, there is a maximal congruence $\t_{a,b}$ of the subalgebra $S=\operatorname{Sg}_A(a,b)$, which separates $a$ and $b$ (i.e. such that $(a,b)\not\in \t_{a,b}$.) Also, there is a ternary polymorphism $g$, which is a majority operation on $\{a/\t_{a,b},b/\t_{a,b}\}$. That is,
$$g(a/\t_{a,b},a/\t_{a,b},b/\t_{a,b})=g(a/\t_{a,b},b/\t_{a,b},a/\t_{a,b})=g(b/\t_{a,b},a/\t_{a,b},a/\t_{a,b})=a/\t_{a,b}.$$
The analogous sequence of equalities holds when the roles of $a/\t_{a,b}$ and $b/\t_{a,b}$ are reversed. However, in that case, there are two proper absorbing subuniverses of $S=\operatorname{Sg}_A(a,b)$, containing $a$ and $b$, respectively. In fact, those subuniverses are $a/\theta_{a,b}$ and $b/\theta_{a,b}$. Recall that, if all the operations of an algebra are idempotent, any congruence class is a subalgebra.

As we have seen in Section~\ref{sec:multisorted}, in the case of a multisorted core, if there is a subuniverse $\mathbb{S}$ of $\A$, which has a proper absorbing subuniverse $\mathbb{S}'<\mathbb{S}$, no solution of $\operatorname{CSP}(\A)$ can pass through an element of $\mathbb{S}'$. Thus, any two vertices which are endpoints of a yellow edge can be excluded from further consideration.

At this point, the solutions can only be endpoints of red and blue edges in the colouring. Next, we will see how to eliminate at least one endpoint from every red edge from further consideration. To do that, we will adopt the conventions and the notation from Section 5.1 of~\cite{bulatov2003conservative}. By Proposition 2 from~\cite{bulatov2003conservative}, there is a fixed, global binary operation $f$ which, for every red edge $\{a,b\}$, witnesses the fact that $\{a/\theta_{a,b},b/\theta_{a,b}\}$ is a red edge, i.e. that $\operatorname{Sg}(a,b)/\theta_{a,b}$ is a simple algebra which has semilattice behaviour. In addition to being commutative (and associative) the operation $f$ also has the property that it is idempotent in both coordinates:
$$f(x,f(x,y))=f(f(x,y),x)=f(x,y).$$
Also, Bulatov's operation $f$ can be chosen in such a way that, if $(a,b)$ is a red edge, $(a,f(a,b))$ is a ``thin" red edge. That is to say, $f$ can be chosen so that the maximal congruence is trivial.

If, for some pair $a,b\in A$, $a\neq b$,
$$f(a,b)=f(b,a)=b$$
we write $a\succ b$ and say that the edge is oriented $a\rightarrow b$. In particular, if $(a,b)$ is a red edge, we will always have $a\rightarrow f(a,b)$. Using a similar argument as in the yellow edge case, if $(a,b)$ is a red edge, there cannot be any solutions passing through $b$, since $b/\t_{a,b}$ will be an absorbing subuniverse for $\operatorname{Sg}_A(a,b)$.

Now, let $\B$ be an arbitrary subalgebra of $\A$, with a maximal congruence $\t$. In that case, $\B/\t$ can be viewed as a connected graph, all of whose edges are of the same colour~\cite{bulatov2003conservative}. We may assume that either all edges are coloured red or all edges are blue, by earlier analysis of what happens in the case of yellow edges. So, we may assume that $\B/\t$ has only red edges, or only blue edges. We will examine the case when all the edges are red.

Since all red edges can be oriented using the operation $f$, we can now look at the strong components of $\operatorname{Gr}(\mathbb{B})$, or, equivalently, we can look at the quasiordering relation on $\B/\t$, denoted $\succ$ as above. Using the same argument as before, there cannot be any solutions in strong components of the directed graph, which are not maximal under the partial order induced by $\succ$. Namely, if $b/\t$ is in a strong component, which is not maximal, then, there will be a directed $\succ$-path from some $a/\t$, which lies in a maximal component (the directed graph is connected). As before $a/\t$ will absorb $b/\t$ and we are in a core structure. 

So, we need only consider the elements in $\succ$-maximal components. First, we observe that if any maximal strongly connected component has more than one element, then, for any $a/\t$ in that component, there exists $b/\t$ such that $b/\t\succ a/\t$ and the absorption argument shows that no solution can pass through $a$. So, the only maximal components that may contain solutions are the ones consisting of a single element. Can there be more than one $\succ$-maximal component all of size $1$? We will show that, in that case, the problem can be reduced to a smaller structure in the given coordinate.

Suppose there is more than one $\succ$-maximal element. Then, for any maximal element $c/\t$, $p(y)=f(c/\t, y)$ is not a permutation. This is due to the fact that, if $c/\t, d/\t$ are two distinct $\succ$-maximal elements, we have $f(c/\t,d/\t)$, cannot be a maximal element, based on the properties of $f$ from~\cite{bulatov2003conservative}. In fact, $f(c/\t,y)$ will not be a permutation for \emph{any} $c/\t$, assuming there are at least two maximal elements. The same holds for $q(x)=f(x,c/\t)$.

By Lemma 9 of~\cite{maroti}, using $f$ as the binary operation $p$, $\B/\t$ can be eliminated from the set of templates, which, essentially means that there is a polynomial time reduction of the domain $\B/\t$ to something properly smaller. For that reason, the only interesting cases of simple algebras $\B/\t$ are the ones having a unique $\succ$-maximal element. Also, we saw that there cannot be any solutions passing through the elements of $\B/\t$ which are not maximal. Thus, in this case, we reduce the search for a solution to a single element, which is $\succ$-maximal.

\subsection{Proof of Correctness}

To complete our proof, we show that passing our $\#\textsc{L}$ consistency test is equivalent to the CSP having a solution. We begin by showing that any inconsistencies can be viewed as certain paths in the multiconsistency graph.

Let $i_0\in I$, where $I$ is the domain (index set) of the CSP instance, and let $\mathbb{C}$ be a subuniverse of $\mathbb{D}_{i_0}$, the $I$-th domain of the instance. Also, let $\theta_C$ be the associated maximal congruence of $\mathbb{C}$. Let $I_C$ be the group of $\mathbb{C}/\theta_C$ and $j\in I_C$. Then, there exists the maximal subuniverse $\mathbb{C}_j$ of $\mathbb{D}_j$ such that the restriction of a binary relation defined by a simple path of the variables in group $I_C$ of $i_0$ with the starting vertex $i_0$ and the ending vertex $j$, to $C\times C_j$ is a subdirect product, along with its maximal congruence $\theta_{C_j}$, such that the graph of the relation induced by the said simple path in $I_C$, restricted to $C/\theta_c\times C_j/\theta_{C_j}$ is the graph of an isomorphism of two simple affine algebras, which are polynomially equivalent to the same Abelian $p$-group $\mathbb{A}$, such that $|A|=p^k$, for some prime $p$ and $k\geq 1$. This follows from the definition of the group of a subuniverse of a domain and the properties of subdirect products of two simple Maltsev algebras.

The restrictions of constraint relations $E_{i,j}$ to all $C_{j}/\theta_{C_j}$, where $j\in I_C$, correspond to isomorphisms between two copies of the $p$-group $\mathbb{A}$ and, for that reason, we can identify the restriction of the instance to $C_{j}/\theta_{C_j}$, where $j\in I_C$, with a system of linear equations, all in two variables, over the Abelian group $\mathbb{A}$. Furthermore, every equation, corresponding to the restriction of the constraint relation $E_{i,j}$ can be viewed as the linear equation $x_j=x_i+a$, where $a\in A$, or the absence of an equation if $E_{i,j}$ corresponds to the full direct product of the two copies of $\mathbb{A}$. 

Now, a simple path from $C_{i'}$ to $C_{j'}$, in the restriction of the multiconsistency graph to the domains $C_{j}/\theta_{C_j}$, all of whose edges correspond to the binary constraints which induce isomorphisms between the two copies of the Abelian group, where $i', j', j\in I_C$, corresponds to combining the equations of the above form using the group addition to obtain an equation of the form $x_{j'}=x_{i'}+a$, for some $a\in A$. If our simple path is closed, starting and ending at the domain with the index $i_0$, we obtain an equation of the form $x_{i_0}=x_{i_0}+a$, for $a\in A$, which can only have a solution if $a=0$, i.e. if the starting and ending points in $C/\theta_C (=\mathbb{A})$, in any such path coincide. Conversely, if the endpoints of any such closed simple path coincide in our restriction of the multiconsistency graph, the solutions of the instance of the CSP (or, equivalently, the corresponding system of linear equations over $\mathbb{A}$) can be chosen consistently.

\begin{lem}
    If a CSP passes the $\#\textsc{L}$ consistency tests then it has a solution.
\end{lem}

    \proof For our base case we assume all domains are simple in which case we can reduce all the domains with only red edges to a point and then solve the CSP over domains containing only blue edges in $\textsc{MOD}_p \textsc{L}$.

    So suppose now that we have a domain $\D_{i_0}$ which is non-simple and we have our associated multiconsistency graph $G$ which passed the consistency checks. Define the group of a domain $\A_i$ to be the indices of the domains which can be reached from $\A_i$ along isomorphism paths. Let $\t_{i_0}$ be the maximal congruence associated with $\D_{i_0}$, let $I_0$ be the group of $\D_{i_0}/\t_{i_0}$, and let $\C$ be any block of $\D_{i_0}/\t_{i_0}$. We now have a $(1,2)$ consistent subinstance $G'$ where the ${i_0}^{th}$ domain is $\C$ and the other domains in the group are the isomorphic copy of $\C$ and the remaining domains are as before. Recall that since the subinstance passes the Datalog check, every pair of domains forms a subdirect product. If this new instance passes our $\#\textsc{L}$ consistency check then the inductive hypothesis applies and we get a solution to $G'$. 

\begin{claim}
    The group of $\C$ is the same as the group of $\D_{i_0}/\t_{i_0}$    
\end{claim}

Certainly the group of $\D_{i_0}/\t_{i_0}$ must be contained in the group of $\C$. So suppose there is an index $j$ in the group of $\C$ which is not in the group of $\D_{i_0}/\t_{i_0}$. This implies that between any domain whose index is in the group and the domain of index $j$ we do not have any possible isomorphism. Thus by Theorem~\ref{maltsev} they are full direct products and so we have that it is still a full direct product when we reduce our domain to $\C$. Thus $j$ cannot be in the group of $\C$ and the claim is proven.

Thus if there is a path which leads to an inconsistency in $G'$ then we must also have such an inconsistency in $G$ as the path would also be included in $G$, so $G'$ must pass the consistency check. Thus we have a solution in $G'$ which will also be a solution in $G$.\qed

\bibliography{bibliography} 
\bibliographystyle{alphaurl}

\end{document}